\documentclass[prb,twocolumn,showpacs,preprintnumbers,amsmath,amssymb,superscriptaddress]{revtex4}

\usepackage{graphicx}
\usepackage{dcolumn}
\usepackage{bm}

\begin{document}

\title{
Electron spin phase relaxation of phosphorus donors
in nuclear spin enriched silicon}
\author{Eisuke Abe}
\affiliation{
Department of Applied Physics and Physico-Informatics,
Keio University, and CREST-JST, 3-14-1 Hiyoshi, Yokohama 223-8522, Japan}
\author{Kohei M. Itoh}
\email{kitoh@appi.keio.ac.jp}
\affiliation{
Department of Applied Physics and Physico-Informatics,
Keio University, and CREST-JST, 3-14-1 Hiyoshi, Yokohama 223-8522, Japan}
\author{Junichi Isoya}
\affiliation{
Research Center for Knowledge Communities, University of Tsukuba,
1-2 Kasuga, Tsukuba City 305-8550, Japan}
\author{Satoshi Yamasaki}
\affiliation{
Diamond Research Center,
National Institute of Advanced Industrial Science and Technology,
Tsukuba Central 2, 1-1-1 Umezono, Tsukuba City 305-8568, Japan}
\date{\today}

\begin{abstract}
We report a pulsed EPR study of
the phase relaxation of electron spins bound to phosphorus donors
in isotopically purified $^{29}$Si
and natural abundance Si ($^{\mathrm{nat}}$Si)
single crystals measured at 8 K.
The two-pulse echo decay curves for both samples
show quadratic dependence on time,
and the electron phase relaxation time $T_M$ for $^{29}$Si
is about an order of magnitude shorter than that for $^{\mathrm{nat}}$Si.
The orientation dependence of $T_M$ demonstrates
that the phase relaxation is caused
by spectral diffusion due to flip-flops of the host nuclear spins.
The electron spin echo envelope modulation effects in $^{29}$Si
are analyzed in the frequency domain.
\end{abstract}
\maketitle

Group-V impurities in silicon have been studied extensively
in semiconductor physics.
Experimental techniques such as infrared absorption, photoluminescence,
and electron paramagnetic resonance (EPR)
have revealed detailed properties of the impurity centers.
EPR is particularly convenient for the identification of defect structures
since the hyperfine (hf) interaction
is a sensitive probe of the spatial distribution of the electron wavefunction.
For instance, Feher and later Hale and Mieher applied
an electron nuclear double resonance (ENDOR) technique to this system,
and measured hf interactions between the donor electron spins and
their neighboring host nuclear spins~\cite{F59,HM69}.
These experimental works,
together with theoretical investigations~\cite{IM75},
have deepened our understanding of shallow donor impurities.

Recently, Kane and others gave a new perspective to the donors in Si,
a playground for solid-state quantum information processing,
since nuclear and electron spins in semiconductors can be regarded
as well-isolated two-level systems: qubits~\cite{K98,VYW+00,LGY+02}.
If the donor electrons are qubits,
$^{29}$Si nuclei that have spin-1/2 and occupy 4.67\%
of the lattice sites in natural Si ($^{\mathrm{nat}}$Si)
are decoherence sources
as their flip-flops produce fluctuations of the local fields.
Indeed, $^{29}$Si-depleted, isotopically controlled $^{28}$Si:P
exhibited the coherence time two orders of magnitude longer
than $^{\mathrm{nat}}$Si:P~\cite{GB58,TLA+03},
demonstrating that such nuclear spin diluted Si
would be indispensable for building a practical Kane-type quantum computer.
On the other hand, a study of the decoherence
caused by the spectral diffusion arising from nuclear flip-flops
requires a material of the opposite class, nuclear spin enriched Si.
This novel material is also interesting 
because of its similarity to III-V materials
in that the electrons are localized in a sea of nuclear spins,
and more preferable for our purpose
owing to the negligibly-small spin-orbit interaction in bulk Si,
which could otherwise contribute to decoherence.

In this Rapid Communication,
we report the phase relaxation time $T_M$ for P donor electron spins
in isotopically purified $^{29}$Si and $^{\mathrm{nat}}$Si measured at 8 K.
The temperature was chosen
so that $T_M$ would not be affected by the spin-flip time $T_1$.
The ground-state electron can be excited by absorbing a phonon
if the phonon energy is comparable to the transition energy
from the A$_1$ ground state to the E or T$_2$ excited states.
When returning to the ground state,
the electron is subject to a spin-flip at a certain probability.
This $T_1$ process, known as an Orbach process, also limits $T_M$
over the temperature range from 10 K to 20 K~\cite{TLA+03,YJK+03}.
While $T_1$ is dominated by the Orbach process down to 6 K,
and extends exponentially with cooling~\cite{C62,C67},
$T_M$ starts to deviate from $T_1$ and becomes
insensitive to the temperature below about 10 K~\cite{TLA+03,YJK+03}.
Since our spin echo experiments require each pulse sequence to be repeated
at time intervals much longer than $T_1$,
we found 8 K to be an appropriate temperature,
low enough for $T_M$ not to be limited by $T_1$
but high enough to ensure a reasonable measuring time.

A Cz-grown single crystal of $^{29}$Si, enriched to 99.23\%,
had a rectangular shape with its long axis in the [1$\bar{1}$0] orientation.
The sample contained 1.8 $\times$ 10$^{15}$ P/cm$^3$
with the compensation of 1.0 $\times$ 10$^{15}$ B/cm$^3$.
Further information on this crystal is provided
in Ref. \onlinecite{IKU+03}.
A $^{\mathrm{nat}}$Si sample was cleaved
from a commercial high-quality wafer containing 0.8 $\times$ 10$^{15}$ P/cm$^3$
with a negligible amount of compensation.
The net donor concentrations of both samples were kept low
so that the dipolar or exchange interactions between donors
would be suppressed~\cite{S55,CH72}.
Pulsed experiments were carried out using a Bruker Elexsys E580 spectrometer,
and samples were kept in an Oxford ER4118CF cryostat.
Temperature was controlled with an Oxford ITC503 temperature controller.
The echo-detected EPR spectra,
in which the intensity of the Hahn echo
was measured as a function of the external magnetic field,
consisted of two Gaussian-shaped lines separated by 4.2 mT.
The splitting is due to the hf interaction with $^{31}$P,
and each line is inhomogeneously broadened
by the surrounding $^{29}$Si nuclei.
The linewidths (FWHM) are 0.26 mT for $^{\mathrm{nat}}$Si
and 1.2 mT for $^{29}$Si.
In the following experiments, the external magnetic field was set to
the center of the line at higher fields ($B_0$ = 348 mT).
$T_1$ was measured using an inversion recovery method
($\pi$-$t$-$\pi$/2-$\tau$-$\pi$-$\tau$-echo),
and is 16 ms for $^{\mathrm{nat}}$Si and 4.4 ms for $^{29}$Si.
As the temperature dependence of the Orbach process is given by
$1/T_1 = R \exp(-\Delta/kT)$,
where $R$ is the rate constant and $\Delta$ the valley-orbit splitting energy,
the difference in $T_1$ between samples could arise in part from
a slight difference in the actual sample temperatures.
The isotope shift of $\Delta$ is unlikely to cause the difference in $T_1$,
since it was not observed in
our infrared photoconductivity measurement on the $^{29}$Si crystal
within the resolution used~\cite{IKU+03}.
The presence of compensation and dislocation (10$^2$ cm$^{-2}$)
in the $^{29}$Si crystal can alter $T_1$,
changing $R$~\cite{S63,YH68}.
However, we can conclude
that this difference in $T_1$ has little effect on the difference in $T_M$
presented below, based on the previous assumption
that $T_1$ does not contribute to $T_M$.

The phase relaxation was investigated
using a two-pulse spin echo method
($\pi$/2-$\tau$-$\pi$-$\tau$-echo,
where the interpulse delay $\tau$ was varied
in 800 ns steps for $^{\mathrm{nat}}$Si
and 40 ns steps for $^{29}$Si.
The $\pi$/2 pulse was 16 ns.).
The samples were rotated around the [1$\bar{1}$0] axis
perpendicular to $\bm{B}_0$.
We define $\theta$ as the angle between $\bm{B}_0$ and [001];
therefore,
$\theta$ = 0$^{\circ}$ when $\bm{B}_0 \parallel$ [001],
$\theta$ = 55$^{\circ}$ when $\bm{B}_0 \parallel$ [111], and
$\theta$ = 90$^{\circ}$ when $\bm{B}_0 \parallel$ [110].
Since the echo-detected EPR spectra were
independent of the crystal orientation
and no other EPR signals were found,
the alignment of the crystal from an EPR signal was not applied here.
We estimate the uncertainty in $\theta$ to be less than 5$^{\circ}$.
\begin{figure}
\includegraphics[scale = 0.4]{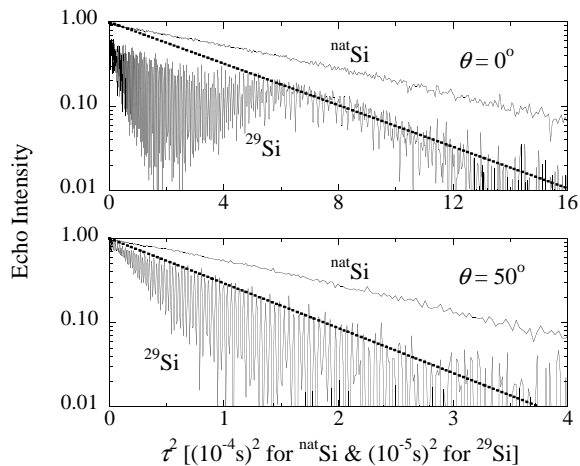}
\caption{\label{echodecay}
The two-pulse electron spin echo decay curves
as a function of $\tau^2$
at $\theta$ = 0$^{\circ}$ and $50^{\circ}$.
Note that the unit of the horizontal axis differs for each sample,
and the scale for each $\theta$.
The dotted lines in $^{29}$Si are the fits to the echo envelope decays.
The oscillation observed in $^{29}$Si is ESEEM. See text.}
\end{figure}
Figure~\ref{echodecay} shows the echo decay curves
at $\theta$ = 0$^{\circ}$ and 50$^{\circ}$.
Although so-called electron spin echo envelope modulation (ESEEM)
obscures the echo envelope decays,
they clearly obey a quadratic decay law, expressed as $\exp(-m \tau^2)$.
A single-exponential term $\exp(-2 b \tau)$ is, if present at all, quite small.
Thus $T_M$ can be defined as the time at which an echo envelope
damps to 1/$e$ of its initial value, i.e., $T_M = 2 m^{-1/2}$.
We note that our temperature setting and assumption on the $T_1$ effect
are justified \textit{posteriori} by the fact
that the echo decay curves are not single-exponential
and that $T_M$ for each sample is much shorter than the respective $T_1$.

\begin{figure}
\includegraphics[scale = 0.4]{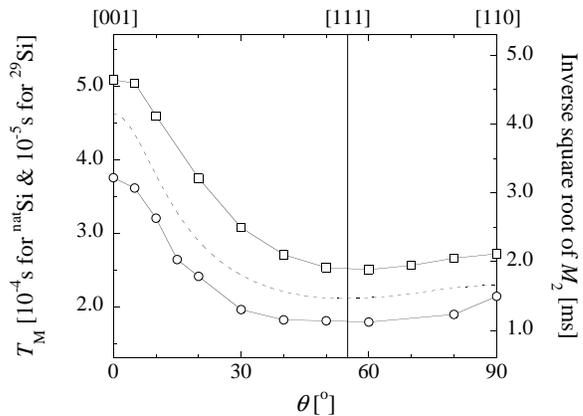}
\caption{\label{tm}
The orientation dependence of $T_M$
for $^{\mathrm{nat}}$Si and $^{29}$Si (left axis).
The open squares ($\square$) represent $T_M$ for $^{\mathrm{nat}}$Si,
and the open circles ($\bigcirc$) for $^{29}$Si.
Note that the unit of the vertical axis differs for each sample.
The inverse square root of $M_2$ calculated based on the method of moment
is also shown by a dashed line (right axis).}
\end{figure}
The orientation dependence of $T_M$ given in Fig.~\ref{tm} shows
$T_M$ to be longest at $\theta$ = 0$^{\circ}$,
and shortest around $\theta$ = 50$^{\circ}$ $\sim$ 60$^{\circ}$~\footnote{
This tendency has also been suggested
by A. M. Tyryshkin and S. A. Lyon experimentally (unpublished).}.
The dependence manifests the fact
that the phase relaxation is caused
by $^{29}$Si nuclei mutually coupled via the dipolar interactions.
This can be verified by calculating the second moment $M_2$
of the $^{29}$Si nuclear spin system.
$M_2$ calculated with Van Vleck's method of moment
is the sum of squared dipolar fields produced by the nuclei~\cite{V48},
and its inverse square root is a convenient measure of the nuclear $T_2$.
$(M_2)^{-1/2}$ for a 100\%
$^{29}$Si crystal is shown in Fig.~\ref{tm} by a dashed line.
Correlations between the electron $T_M$ and the nuclear $M_2$ are apparent.
As $M_2$ directly reflects the strength of the nuclear dipolar couplings,
its orientation dependence is understood qualitatively as follows:
When $\bm{B}_0$ is along [111],
one of the four nearest-neighbor bonds of the Si atoms
is parallel to $\bm{B}_0$,
and this pair of nuclei gives rise to the strongest coupling;
hence, $M_2$ takes its maximum.
With $\bm{B}_0$ along [001], all the dipolar couplings
between nearest neighbors are frozen
since the angle between $\bm{B}_0$ and the vector
connecting the nearest neighbors is a so-called magic angle;
hence, $M_2$ takes its minimum.
In fact such an experimental $T_2$ has been reported
for NMR of $^{13}$C diamond,
a material similar to $^{29}$Si~\cite{SSS95}.
As the lineshape studies of NMR spectra for $^{13}$C diamond
and $^{29}$Si have revealed
that they share essentially the same line-broadening mechanism,
$T_2$ for $^{29}$Si will show the same tendency as
that for $^{13}$C diamond if measured~\cite{LBP+94,VMY+03}.

Although the comparison with $M_2$ works qualitatively,
it provides little information on the actual value of $T_M$.
Theoretical estimation of $T_M$ must take
the hf interaction between the electron and host nuclei into account
as well as the nuclear dipolar coupling.
Generally, to characterize a system where the electron phase relaxation
is caused by the spectral diffusion
due to flip-flops of the host nuclear spins,
the diffusion barrier that prevents the flip-flops within its bounds
must be considered~\cite{ST79}.
As the Fermi contact hf interaction,
which is proportional to the density of the electron wavefunction
$|\Psi(\bm{r}_i)|^2$, varies from site to site,
a flip-flop of a certain pair of nuclei occurs only when
the difference of the hf interaction between the pair
is small enough to satisfy the condition of energy conservation.
The condition must be evaluated for each pair,
since $|\Psi(\bm{r}_i)|^2$ does not
decrease monotonically with increasing $r_i$
but oscillates due to the multi-valley nature of Si.
Such a theoretical treatment
has been proposed by de Sousa and Das Sarma~\cite{SS03},
it is therefore interesting to compare our results with theirs~\footnote{
Theoretical values given here were provided by R. de Sousa
(private communication).}
Theory predicts the observed angular dependence correctly,
but overestimates $T_M $ by about a factor of 3 for both samples.
This already-reasonable agreement becomes even better
if we take the ratio of $T_M$ between the samples.
Indeed, the theoretical ratio of $T_M$ for $^{\mathrm{nat}}$Si
to that for $^{29}$Si falls between 11.2 and 11.8,
while the experimental ratio lies between 11.2 and 14.4.
Given the difficulty in determining the precise $T_M$ due to ESEEM,
their calculation is in good agreement with our experiments.
Another comparison is to take the ratio of
the maximum ($\theta$ = 0$^{\circ}$)
and minimum ($\theta$ $\sim$ 55$^{\circ}$) $T_M$.
Theory yields 2.7 for $^{\mathrm{nat}}$Si and 2.9 for $^{29}$Si,
compared with experimental values of
2.0 for $^{\mathrm{nat}}$Si and 2.1 for $^{29}$Si.
The larger values in the theory
may indicate the presence of an isotropic contribution to $T_M$,
but it is not clear at this stage
whether other decoherence mechanisms must be incorporated
or an improved theory of nuclear-induced spectral diffusion suffices
to explain the discrepancies revealed here.
Incorporating non-Markovian nuclear flip-flop processes
would certainly be an interesting refinement of the theory,
while the stochastic treatment proved valid even for $^{29}$Si.
It is also noteworthy that recent multiple-pulse NMR studies in Si
provide a glimpse into the complicated behavior
of this seemingly-simple dipolar coupled system~\cite{WS03, DLM+03, LMY+03}.
Clearly, more experimental and theoretical investigation
is necessary for a full understanding of the phenomena.

We now turn our attention to the remarkable feature of the decay curves: ESEEM.
The origin of ESEEM can be described briefly as follows:
If the nuclear spin feels, in addition to the external magnetic field,
the moderate hf field produced by the electron spin,
the nuclear spin precesses around an effective magnetic field
which is tilted from the external magnetic field,
i.e., $m_I$ is no longer a good quantum number.
Due to this state mixing, formally forbidden nuclear-spin-flip transitions 
($\Delta m_S = \pm 1, \Delta m_I = \pm 1$) can occur,
and interfere with allowed transitions
to produce beats in the electron spin echo envelope.
In two-pulse experiments for an $S = 1/2$, $I = 1/2$ spin system,
the modulation contains the ENDOR frequencies $\nu_{+}$ and $\nu_{-}$,
and their sum and difference $\nu_{+} \pm \nu_{-}$.
When many nuclei are coupled to the same electron spin,
some combination frequencies are also contained
since the two-pulse ESEEM is the product of individual modulation functions.

We analyzed the ESEEM spectra in the frequency domain.
Although ESEEM was also observed in $^{\mathrm{nat}}$Si,
we treat only the case of $^{29}$Si here
because the larger modulation depth in $^{29}$Si facilitated the analysis.
Also, the modulation depth is strongly angular dependent
(Fig.~\ref{echodecay}),
since the degree of state mixing depends on
both the position of each nuclear spin
and the orientation of the external magnetic field.
To obtain a frequency domain spectrum,
the slowly decaying part of a time domain spectrum was subtracted first,
then the remaining modulation was Fourier-transformed~\footnote{
In the present experiments, a step of $\tau$ was set to 40 ns,
hence the Nyquist frequency is 12.5 MHz.
As the data were taken from $\tau$ = 320 ns,
all the modulation components that decayed within 320 ns
cannot be recovered in the frequency domain spectra.}.
\begin{figure}
\includegraphics[scale = 0.4]{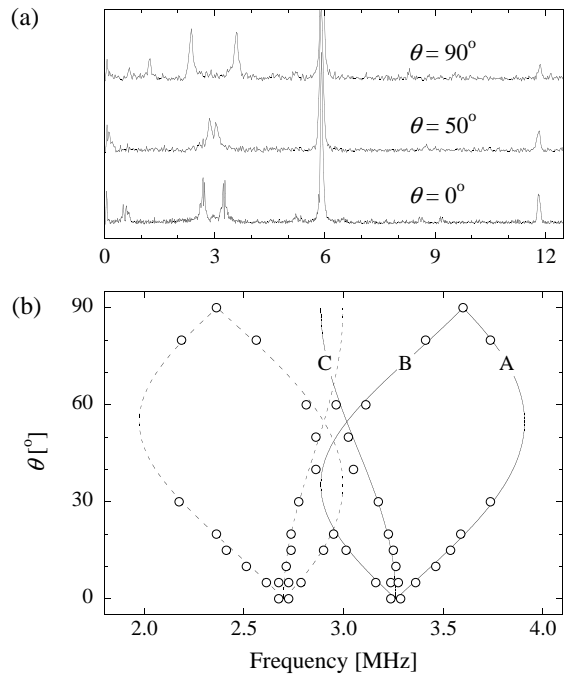}
\caption{\label{fteseem}
(a) The frequency domain ESEEM spectra in $^{29}$Si
at $\theta$ = 0$^{\circ}$, 50$^{\circ}$, and 90$^{\circ}$.
The vertical axis is shown in an arbitrary unit and shifted for clarity.
(b) The orientation dependence of the ENDOR frequencies around 3 MHz.
They show a $\langle$111$\rangle$-axis pattern.
The solid (dashed) lines are for $\nu_{+}$ ($\nu_{-}$).
Which lattice site produces lines A, B, and C is explained in the text.}
\end{figure}
Figure~\ref{fteseem}(a) shows the frequency domain spectra
at $\theta$ = 0$^{\circ}$, 50$^{\circ}$, and 90$^{\circ}$.
Peaks around 3 MHz are the ENDOR lines,
and their angular dependence is shown in Fig.~\ref{fteseem}(b),
from which we see a $\langle$111$\rangle$-axis pattern~\cite{HM69}.
The ENDOR frequencies
for an axially symmetric hf tensor with an isotropic g-factor
are given by~\cite{AJ01}
\[\nu_{\pm} = \sqrt{
  \left( \nu_I \pm
  \frac{a_{\mathrm{iso}} + b (3 \cos^2 \varphi_i -1)}{2} \right)^2 +
  \left( \frac{3 b \sin 2 \varphi_i}{4} \right)^2},
\]
where $\nu_I$ is the nuclear Larmor frequency,
$a_{\mathrm{iso}}$ the isotropic hf coupling constant,
$b$ the anisotropic hf coupling constant,
$\varphi_i$ the angle between $\bm{B}_0$ and the unique axis of the hf tensor.
$\nu_I$ is calculated to be 2.94 MHz
as the gyromagnetic ratio of $^{29}$Si nuclei is 8.46 MHz/T.
$\nu_{\pm}$ calculated with
$a_{\mathrm{iso}}$ = 570 kHz and
$b$ = 681 kHz agree well with the experimental results,
as shown in Fig.~\ref{fteseem}(b).
In comparison with hf constants obtained
from previous cw ENDOR experiments~\cite{HM69},
the observed peaks are assigned to shell E (111),
i.e., four nearest neighbors of the donor.
Lines A and B originate from (111) and ($\bar{1}$$\bar{1}$1) sites,
respectively.
Line C is doubly degenerate,
since (1$\bar{1}$$\bar{1}$) and ($\bar{1}$1$\bar{1}$) sites
locate each other at plane symmetric positions
with respect to the (1$\bar{1}$0) plane.
The experimental data corresponding to line C
at $\theta$ = 0$^{\circ}$ and 10$^{\circ}$ split, however.
This suggests that the sample was not exactly rotated
around the [1$\bar{1}$0] axis but slightly off-axis,
most likely due to a small miscut of the crystal.
This assumption is supported by the fact
that the ESEEM in $^{\mathrm{nat}}$Si at $\theta$ = 0$^{\circ}$
did not split (not shown).
The strong peak at 5.9 MHz is the sum frequency,
but signals from shell A (004) are overlapped.
The fourth harmonic is also observed at 11.8 MHz,
and the third harmonic is barely visible around 9 MHz.
We did not observe the third and fourth harmonics in $^{\mathrm{nat}}$Si.
We also observed tiny peaks around 5.2 MHz throughout the angles tested.
They are assigned to shell B (440),
but the detailed angular dependence was untraceable.
A three-pulse stimulated echo method would be suitable for
a more detailed ESEEM study.
From the viewpoint of quantum computing,
ESEEM clearly leads to quantum-gate errors.
For this purpose, time domain analysis
is highly desirable as recently simulated by Saikin and Fedichkin~\cite{SF03}.

In conclusion, we have measured the phase relaxation time $T_M$
of P donor electron spins for $^{\mathrm{nat}}$Si and $^{29}$Si at 8 K.
$T_M$ for $^{29}$Si is an order of magnitude shorter
than that for $^{\mathrm{nat}}$Si
due to much more frequent flip-flops of the host nuclear spins.
The orientation dependence of $T_M$ agrees qualitatively
with $(M_2)^{-1/2}$ for a 100\%
$^{29}$Si crystal calculated with the method of moment,
and quantitatively with the theory of de Sousa and Das Sarma.
Frequency domain analysis revealed
that ESEEM effects originate mainly from the hf interactions
between the donor electron and its nearest neighbor nuclei,
as suggested by Saikin and Fedichkin.
Our results also provide insights into the localized electrons
in III-V materials, such as GaAs,
whose lattice sites are full of nuclei with non-zero spin.
Their phase relaxation would be severely controlled by
nuclear-induced spectral diffusion,
therefore the experimental conditions must be arranged carefully
so that the effects of nuclear spins may be suppressed,
e.g., high magnetic fields, decoupling pulses, etc.
In the near future we plan to prepare a series of samples
with different $^{29}$Si isotopic composition.
Such samples will allow us to carry out systematic relaxation time studies
of the electron and nuclear spins as a function of $^{29}$Si concentration.

We thank H. -J. Pohl for the $^{29}$Si crystal and
R. de Sousa for valuable comments
and kindly providing
his calculation results.
This work was supported in part by the
Grant-in-Aid for Scientific Research No. 64076215.

\bibliography{esr29si}
\end{document}